\documentclass[preprint,12pt]{elsarticle}

\usepackage{amssymb}
 \usepackage{amsthm}
 \usepackage{amsmath}

\usepackage[utf8]{inputenc}
\usepackage[caption=false]{subfig}
\usepackage{graphicx}
\usepackage[hidelinks]{hyperref}

\journal{Commun. Nonlinear Sci. Numer. Simul.}

\begin{document}

\begin{frontmatter}



\title{A test for fractal boundaries based on the basin entropy}

\author{Andreu Puy}
\address{Nonlinear Dynamics, Chaos and Complex Systems Group, Departamento de  F\'isica, Universidad Rey Juan Carlos\\ Tulip\'an s/n, 28933 M\'ostoles, Madrid, Spain}

\author{Alvar~Daza}
\address{Nonlinear Dynamics, Chaos and Complex Systems Group, Departamento de  F\'isica, Universidad Rey Juan Carlos\\ Tulip\'an s/n, 28933 M\'ostoles, Madrid, Spain}
\address{Department of Physics, Harvard University, Cambridge, MA 02138, U.S.A.}

\author{Alexandre~Wagemakers}
\address{Nonlinear Dynamics, Chaos and Complex Systems Group, Departamento de  F\'isica, Universidad Rey Juan Carlos\\ Tulip\'an s/n, 28933 M\'ostoles, Madrid, Spain}

\author{Miguel A.F. Sanju\'{a}n}
\address{Nonlinear Dynamics, Chaos and Complex Systems Group, Departamento de  F\'isica, Universidad Rey Juan Carlos\\ Tulip\'an s/n, 28933 M\'ostoles, Madrid, Spain}
\address{Department of Applied Informatics, Kaunas University of Technology, Studentu 50-415, Kaunas LT-51368, Lithuania}
\date{\today}

\begin{abstract}
In dynamical systems, basins of attraction connect a given set of initial conditions in phase space to their asymptotic states. The basin entropy and related tools quantify the unpredictability in the final state of a system when there is an initial perturbation or uncertainty in the initial state. Based on the basin entropy, the $\ln 2$ criterion allows for efficient testing of fractal basin boundaries at a fixed resolution. Here, we extend this criterion into a new test with improved sensitivity that we call the \textit{$S_{bb}$ fractality test}. Using the same single scale information, the $S_{bb}$ fractality test allows for the detection of fractal boundaries in many more cases than the $\ln 2$ criterion. The new test is illustrated with the paradigmatic driven Duffing oscillator, and the results are compared with the classical approach given by the uncertainty exponent. We believe that this work can prove particularly useful to study both high-dimensional systems and experimental basins of attraction.
\end{abstract}




\end{frontmatter}



\section{\label{sec:Introduction}Introduction}
Dynamical systems often show multistability with the coexistence of several asymptotic states, known as \emph{attractors}~\cite{feudel2008complex}. In dissipative systems, the set of initial conditions that asymptotically approach an attractor is called the \emph{basin of attraction}~\cite{nusse1996basinsSci}. In open Hamiltonian systems, instead of attractors we have exits, and consequently the initial conditions leaving the system by these exits are their \emph{escape basins}~\cite{aguirre2001wada}. In both cases, basin boundaries can either be smooth or fractal curves. Under repeated enlargement, fractal boundaries reveal new structures at arbitrarily small scales. This leads to non-integer dimensions and it is often considered as one of the hallmarks of chaos~\cite{mcdonald1985fractal,aguirre2009fractal}. The existence of fractal basin boundaries is profoundly intertwined with the unpredictability under uncertainty in the initial state of the orbit's attractor; this is the facet of chaos we aim to study here.

The classical method for studying the lack of predictability of a multistable system is via the \emph{uncertainty exponent} $\alpha$~\cite{grebogi1983final}. Basically, it measures the fractal dimension of the boundaries counting the number of boxes that lie between basins at different scales. The uncertainty exponent ranges from zero for the most unpredictable basins to one for smooth boundaries. Nonetheless, the uncertainty exponent presents some unavoidable numerical difficulties, such as accessing arbitrarily small scales or exploring multiple box sizes. Besides, it makes a poor use of the information obtained by sampling the phase space with the boxes, since it only classifies them as certain (lying in the interior) or uncertain (lying on a boundary).

Other tools able to quantify the attractors unpredictability given a finite uncertainty in the initial states are the \emph{basin entropy} and the \emph{boundary basin entropy}~\cite{daza2016basin}. Starting with a similar tiling of phase space, the idea is to compute the probabilities of going to each attractor within a box and exploit Shannon's information entropy. Since their formulation, the basin entropy and the boundary basin entropy have been applied to experiments with cold atoms~\cite{daza2017chaotic}, chaotic scattering~\cite{bernal2018uncertainty, nieto2018resonant}, biological systems~\cite{donepudi2018collective, mugnaine2019basin}, electronic micro/nanodevices~\cite{gusso2019nonlinear}, oscillators~\cite{kong2019special} and astrophysical models~\cite{zotos2018basins, zotos2018newton, zotos2018henon}, among others.

Based on the boundary basin entropy, the \emph{$\ln 2$ criterion}~\cite{daza2016basin} to detect fractal boundaries at a fixed resolution was developed. The use of a single scale to test for fractality makes it both computationally and experimentally convenient. The $\ln 2$ criterion applies to boundaries separating more than two different basins, like the Wada basins~\cite{kennedy1991basins}. The goal of this paper is to extend the $\ln 2$ criterion, presenting a test as a criterion to ascertain for fractal structures with an improved sensitivity that also applies for boundaries separating only two basins.

The article is organized as follows: we start in section~\ref{sec:Fractal structures: a perspective based on the basin entropy} with a quick revision of the basin entropy and its relation to fractal structures on phase space. We continue in section~\ref{sec:Testing for fractal structures with the basin entropy} introducing the fractal test and developing it for the case when basins are known exactly in two-dimensional phase spaces. Then, in sec.~\ref{sec:Effects of finite grids}, we study finite resolution effects in discrete phase spaces given by finite grids. In sec.~\ref{sec:Example}, we give a recipe of the test and illustrate it with an example. Moreover, in sec.~\ref{sec:Extension any dimension}, we extend the test to phase spaces in any dimension. Finally, we summarize and discuss the results in sec.~\ref{sec:conclusions}.

\section{Basin entropy and the definition of fractality\label{sec:Fractal structures: a perspective based on the basin entropy}}

Since our main goal is the identification of fractal boundaries, we first need to consider the definition of \emph{fractal}. Even though fractals are not defined in the literature in a precise and unambiguous manner, they typically share some of the following properties~\cite{falconer2004fractal}: (1) fine or detailed structure at arbitrarily small scales, (2) local and global irregularity non-describable by ordinary geometry, (3) some notion of self-similarity, (4) a `fractal dimension' greater than the topological dimension and (5) simple and perhaps recursive definitions. Here, we refer to fractal structures as those fulfilling at least properties (1), (2) and (4). Nonetheless, numerically we are always able to reach only up to a given scale. This connects with the experience that physical systems do not exhibit `true' fractal structures, in the sense that there are only some finite accessible or even defined scales~\cite{avnir1998geometry}.

\begin{figure}
  \centering
    \subfloat[\label{fig:duf smooth}]{\includegraphics[width=0.32\textwidth]{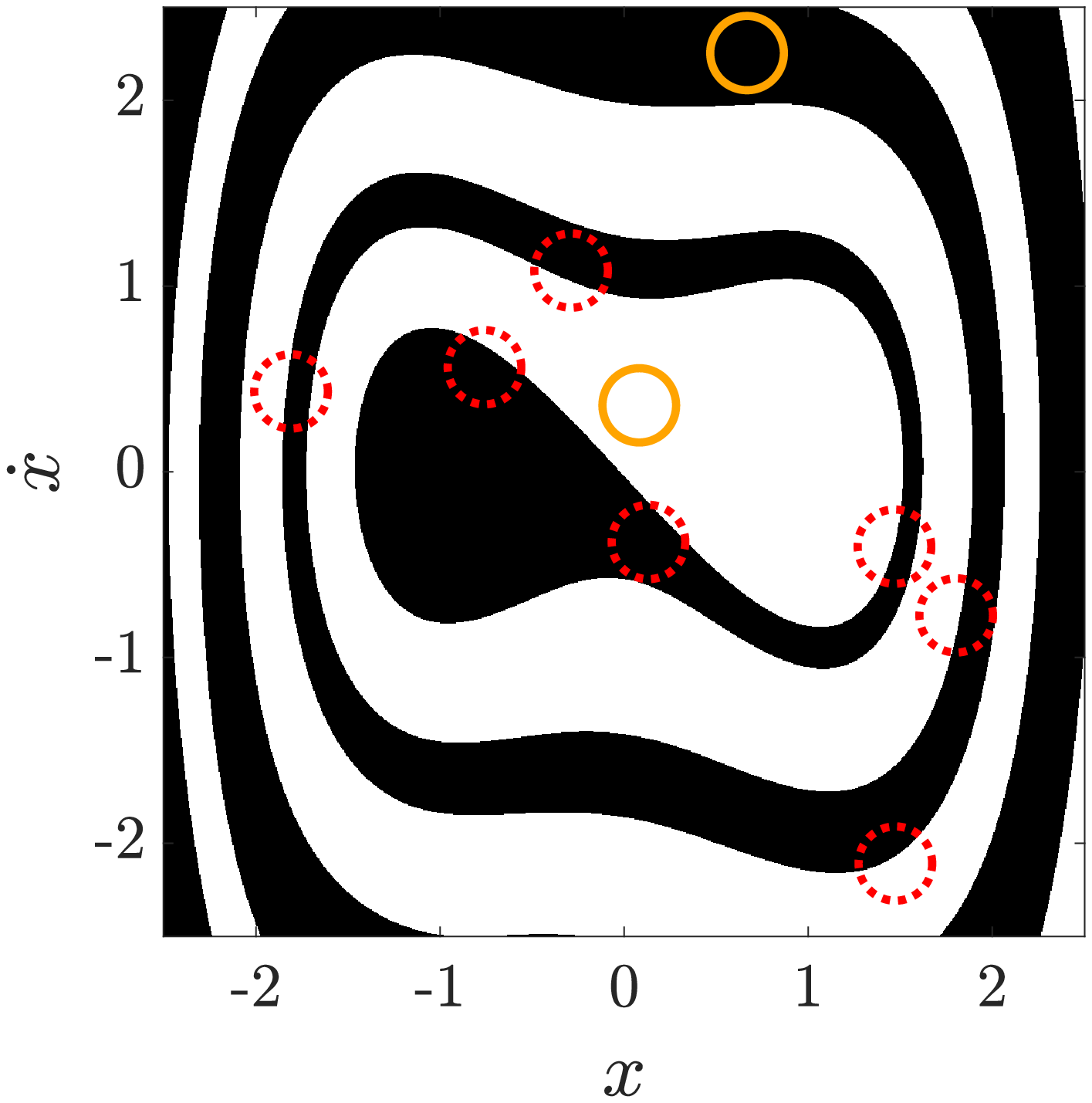}}
    \hspace{0.01\textwidth}
    \subfloat[\label{fig:duf fractal 2 basins}]
    {\includegraphics[width=0.32\textwidth]{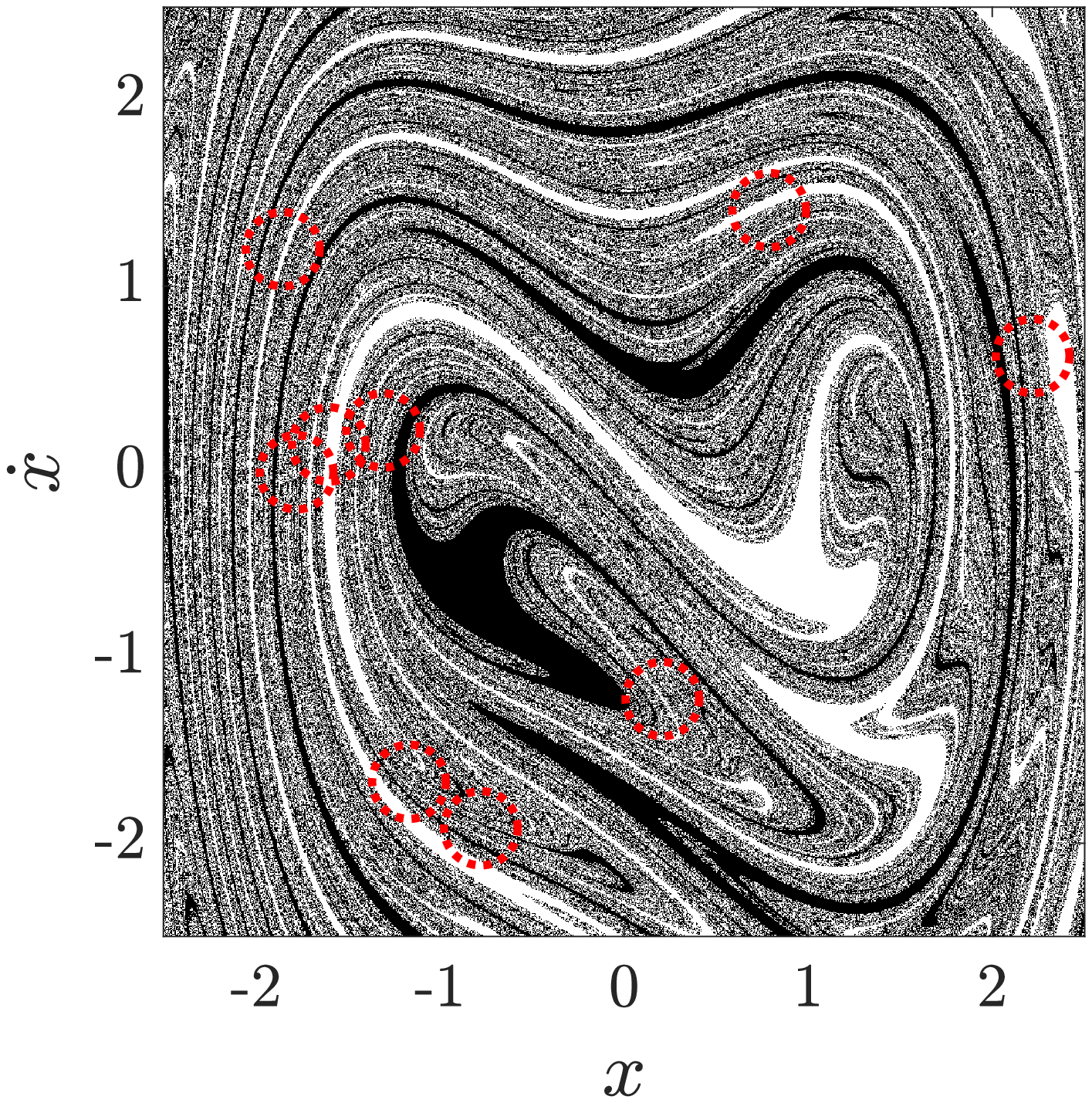}}
    \hspace{0.01\textwidth}
        \subfloat[\label{fig:duf fractal multiple basins}]
    {\includegraphics[width=0.32\textwidth]{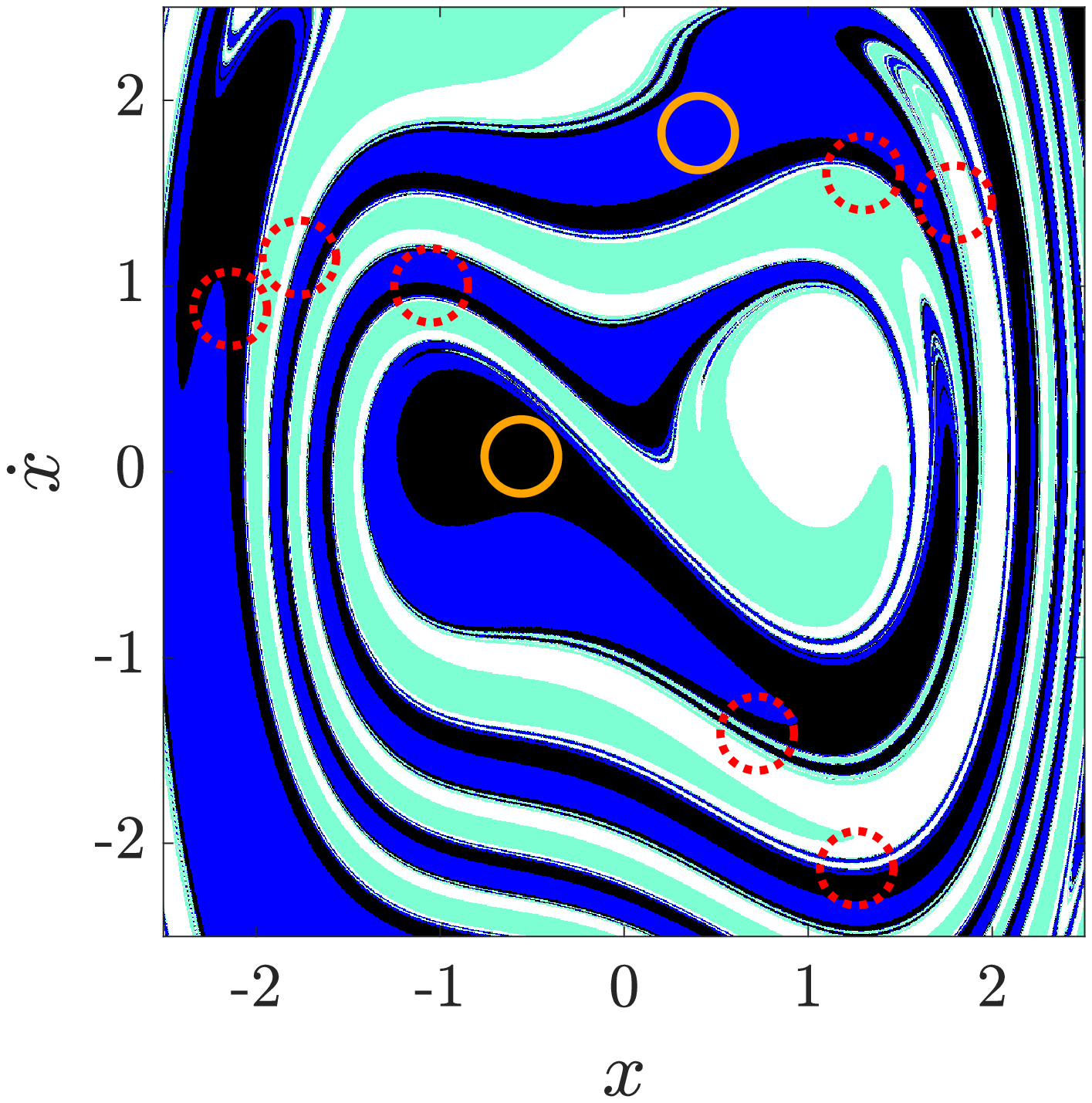}}

  \caption{To compute the basin entropy, we draw boxes of size $\varepsilon$ from the phase space. For each box, we calculate the information entropy from the fractions of each basin within the box; then, averaging over all boxes we calculate the basin entropy $S_b$; and averaging only over boundary boxes (red dashed circles), the boundary basin entropy $S_{bb}$. These basins are from the periodically driven Duffing oscillator $\ddot{x} + 0.15 \dot{x} - x + x^3 = F \sin \omega t$, and (a) $F=0.100$, $\omega = 0.200$, (b) $F=0.395$, $\omega = 1.617$ and (c) $F=0.128$, $\omega = 1.106$. For $N_b = 10^4$ disks boxes in the boundary of radius $\varepsilon = 0.025$, the figures are ordered from left to right by the boundary basin entropy: $S_{bb} =  0.465 \pm 0.002 < 0.6323 \pm 0.0011 \leq \ln 2 < 0.760 \pm 0.004$. Using the $\ln 2$ criterion, only basins (c) are tested with fractal boundaries.\label{fig:duf basins}}
  \end{figure}

The basin entropy provides a way to characterize phase space structures at a given scale. We consider a region of the phase space with $M$ distinct basins and we sample it using $N$ boxes of size $\varepsilon$ (see Fig.~\ref{fig:duf basins}). The box size $\varepsilon$ dictates the scale to study the boundary structures. These boxes can be thought as initial states either with uncertainty $\varepsilon$ or an initial perturbation $\varepsilon$. We can construct the discrete probability distribution of having a basin $k$ for each box, $p(k)$, as the ratio of the volume of the basin $k$ in the box, $v(k)$,  to the phase space volume comprised by the box, $V$:
\begin{equation}
    p(k) \equiv \frac{v(k)}{V},\label{eq:box prob as volume ratios}
\end{equation}
such that $\sum_{k=1}^M p(k ) = 1$. Its associated information entropy $s$,
\begin{equation}
    s \equiv s(p(1), \dots, p(M)) = \sum_{k=1}^M -p(k  ) \ln{p(k)},\label{eq:definition entropy s}
\end{equation}
measures the final state unpredictability of an initial condition chosen at random within the box defined by $\varepsilon$. It increases monotonically with the number of basins within the box and tends to a maximum value $s = \ln M$ as the probabilities $p(k)$ tend to the equiprobable conditions $p(k) = \frac{1}{M}$ for all $k$.

Then, the \emph{basin entropy} $S_b$ is defined as the average of the entropy of the box $s$ for the total number of boxes $N$,
\begin{equation}
    S_b \equiv \frac{1}{N} \sum_{i=1}^{N} s(i),\label{eq:basin entropy}
\end{equation}
where $i$ labels the boxes. Therefore, for an initial random position in the phase space region that is uncertain within a volume of size $\varepsilon$, the basin entropy quantifies the unpredictability of the orbit's attractor. Moreover, the \emph{boundary basin entropy} $ S_{bb}$ is defined as the average of the entropies $s(i)$ restricted only to the $N_b$ boxes falling on basin boundaries (in Fig.~\ref{fig:duf basins}, these correspond to the red dashed boxes):
\begin{equation}
    S_{bb} \equiv \frac{1}{N_b} \sum_{i=1}^{N_b} s(i). \label{eq:boundary basin entropy}
\end{equation}
The boundary basin entropy quantifies the unpredictability focusing only on the unpredictable regions of the phase space: the basin boundaries.

Based on the boundary basin entropy, the $\ln 2$ criterion provides a sufficient condition to test for fractal boundaries. It is based on the fact that smooth boundaries separate only two basins, with the possible exception of a countable number of points that can separate three or more basins at a time. Therefore, for a sufficient large number of small boxes, for smooth boundaries $S_{bb} \in [0, \ln 2\simeq 0.693]$. This is the case of the basins in Fig.~\ref{fig:duf smooth}, where for a box scale $\varepsilon = 0.025$, $S_{bb} = 0.465 \pm 0.002$. Equivalently, if $S_{bb}$ is significantly larger than $\ln 2$, this implies the boundary will not be smooth (i.e., it will be fractal). This occurs for the basins in Fig.~\ref{fig:duf fractal multiple basins}, where $S_{bb} = 0.760 \pm 0.004$. Nonetheless, the criterion fails for the case of Fig.~\ref{fig:duf fractal 2 basins} with a manifested fractal boundaries between two basins and $S_{bb} = 0.6323 \pm 0.0011$. Following this idea, there is room to make a better test using a single scale with the basin entropy.

\section{\label{sec:Testing for fractal structures with the basin entropy} Beyond the ln 2 criterion: the $S_{bb}$ fractality test}
Simple smooth boundaries are those which are locally flat (as a straight line in 2D). According to our previous definition of a fractal, we consider as a fractal boundary any boundary that is not simple smooth. A simple smooth boundary could also include a few finite points where boundaries from several basins intersect. Nevertheless, in the infinitely fine scale these regions become negligible in front of the other locally flat regions.

Based on this idea and the boundary basin entropy, we can define a statistical test to identify fractal structures, which we name \textit{$S_{bb}$ fractality test}. It consists on comparing the value of the boundary basin entropy $S_{bb}$ of the boundary under study to the theoretical value $S_{bb}$ of a flat boundary, at a given small box scale $\varepsilon$. If they are deemed statistically significantly different, the boundary has a fractal structure. We will define it formally after studying the boundary basin entropy of a flat boundary. Since the boundary basin entropy is defined as an average of the box entropies $s$ and different configuration of the boxes' entropies can give rise to the same value, this test is a sufficient but not necessary condition. Nonetheless, it is more restrictive than the $\ln 2$ criterion and consequently it can detect fractal boundaries in many more cases, e.g. in regions with only two basins.

Assuming a perfect knowledge of the phase space structure for a flat boundary in a two-dimensional phase space, we can derive the theoretical value of the boundary basin entropy. We start by choosing a disk as box shape, because its rotational symmetry under any angle allows to compute the boundary basin entropy independently of the box orientation. Now, having disk boxes in a phase space with a flat boundary, we can obtain the boundary basin entropy by sliding the box from side to side along the boundary. If a disk box has a radius $\varepsilon$ and is centered at the coordinate $x_0 \in  [-\varepsilon, \varepsilon]$ from the perpendicular direction to the boundary with origin on the boundary, the probability $p(x_0)$ to have a disk point in one of the basins is the fraction of the disk given by the circular segment
\begin{equation}
p(x_0) = \frac{1}{2} + \frac{1}{\pi} \left[ \frac{x_0}{\varepsilon}\sqrt{1-\left( \frac{x_0}{\varepsilon} \right)^2} + \arcsin{\frac{x_0}{\varepsilon}}\right].
\end{equation}

Then, $S_{bb}$ is given numerically by:
\begin{equation}
    S_{bb} = \frac{1}{2 \varepsilon } \int_{-\varepsilon}^{\varepsilon} dx_0 \ s\left(p \left(x_0\right), 1 - p \left(x_0\right)   \right) =  0.4395093(6).\label{eq:sb sliding box}
\end{equation}
This means that if we were able to compute exactly the boundary basin entropy of a smooth boundary using small disks we would get that result, and any other number would correspond to a fractal case. However, in practice we always have some errors induced by the use of a finite number of trajectories. The effects that this can introduce into our test are explored in the next section.

\section{\label{sec:Effects of finite grids}Effects of finite grids}

Both numerically and in experiments, basins are commonly calculated using a finite number of grid points. The number of grid points within a box determines the observed probabilities $\hat{p}(k)$ of the box. Ultimately, this lack of detail results in an observed boundary basin entropy $\hat{S}_{bb}$ with a systematic error or bias $\delta_{\hat{S}_{bb}}$. In addition, we are generally limited to draw a finite number of boxes too. This results in an observed boundary basin entropy with a statistical error $\sigma_{\hat{S}_{bb}}$. Here, we study these effects for a two-dimensional square grid with a flat boundary and disk boxes, which provides confidence intervals for our fractal basin boundary test.

\begin{figure}[h]
  \centering
       \includegraphics[width=0.45\textwidth]{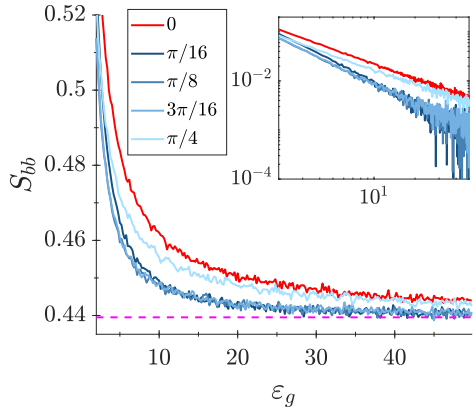}
  \caption{For a grid with a flat boundary and disk boxes of radius $\varepsilon_g$ in grid units, the systematic error of the observed boundary basin entropy $\hat{S}_{bb}$ depends on the angle between the grid and the boundary (legend): (outset) $\hat{S}_{bb}$ in function of  $\varepsilon_g$ and the exact  $S_{bb}$ value (magenta dashed line); (inset) absolute errors of $\hat{S}_{bb}$ in function of $\varepsilon_g$. We considered large grid extensions of around $128 \varepsilon_g$ points per axis and $N_b = 10^5$ disk boxes in the boundary. The worst systematic error is given by the angle $0$ in red. In the infinite grid resolution ($\varepsilon_g \to \infty$) and for any angle $\alpha$, we recover the exact boundary basin entropy $S_{bb}$.\label{fig:grid angular dependency}}
  \end{figure}

The systematic error of the observed boundary basin entropy $\hat{S}_{bb}$ depends on two factors: the extension of the grid and the angular orientation of the grid. The effect of the grid extension decreases with the number of grid points per axis. Our numerical simulations showed that this factor is no longer relevant for larger values than $50 \varepsilon_g$ grid points per axis, where $\varepsilon_g$ is the disk radius in \textit{grid units}. The second factor, the angle between the grid and the boundary, is inherently unavoidable. We have investigated this effect in Fig.~\ref{fig:grid angular dependency}, representing in function of the disk radius $\varepsilon_g$ in grid units and for largely extended grids, $\hat{S}_{bb}$ for different angles (legend) and the exact $S_{bb}$ (magenta dashed line). The figure inset displays the absolute errors of $\hat{S}_{bb}$. Even in the worst case scenario (red line), the systematic error decreases approximately inversely proportional to the disk boxes radius $\varepsilon_g$. Indeed, a power law fit gives an upper bound for the systematic error:
\begin{equation}
\delta_{\hat{S}_{bb}}^\text{UB} \simeq A\varepsilon_g^B,\label{eq:Sbb UB systematic error}
\end{equation}
with $A=0.224 \pm 0.010$ and $B = -1.006 \pm 0.014$.

On the other hand, there is a statistical error of the observed boundary basin entropy $\hat{S}_{bb}$, which is due to the finite number $N_b$ of boxes in the boundary. Indeed, $\hat{S}_{bb}$ is the average of the boxes' entropies in the boundary. This means that, by the central limit theorem, the statistical error follows a Gaussian distribution with a standard deviation given by:
\begin{equation}
\sigma_{\hat{S}_{bb}} = a N_b^{-\frac{1}{2}},\label{eq: Sb statistical error}
\end{equation}
where $a \equiv \sqrt{\frac{1}{N_b - 1 } \sum_{i=1}^{N_b} \left (\hat{S}_{bb} - \hat{s}(i) \right )^2 }$ is the sampling standard deviation and $\hat{s}(i)$ the observed entropy for the box $i$.

Taking into account all the finite grid effects for 2D phase spaces, we can formulate our test for fractal boundaries as follows: \textit{under an infinitesimal disk box with radius $\varepsilon_g$ in grid units and a finite grid that is largely extended (with at least $50 \varepsilon_g$ grid points per axis), if the observed boundary basin entropy $\hat{S}_{bb}$ with a standard deviation $\sigma_{\hat{S}_{bb}}$ (Eq.~\ref{eq: Sb statistical error}) is deemed statistically significant away from a flat boundary, either below the exact boundary basin entropy value $S_{bb}$ (Eq.~\ref{eq:sb sliding box}) or above the upper-bound systematic error value $S_{bb} + \delta_{\hat{S}_{bb}}^\text{UB}$ ($\delta_{\hat{S}_{bb}}^\text{UB}$ is given by Eq.~\ref{eq:Sbb UB systematic error}), the boundary has a fractal structure.} Using one standard deviation $\sigma_{\hat{S}_{bb}}$ for the statistical error, we can express the $S_{bb}$ fractality test by the following sufficient conditions:
\begin{align}
 \hat{S}_{bb}  &< S_{bb} - \sigma_{\hat{S}_{bb}},\label{eq:1st-order fractal test eq 1} \\
\hat{S}_{bb}  &> S_{bb} + \delta_{\hat{S}_{bb}}^\text{UB} + \sigma_{\hat{S}_{bb}}.\label{eq:1st-order fractal test eq 2}
\end{align}

\section{\label{sec:Example}Example of application}
The periodically driven Duffing oscillator is a paradigmatic model that accounts for nonlinear elastic effects in large displacements of a forced damped elastic structure. It is defined by:
\begin{equation}
\ddot{x} + \gamma \dot{x} - x + x^3 = F \sin \omega t,
\end{equation}
where $x$ is the displacement of the oscillator at time $t$, $\gamma$ is the damping coefficient, $F$ is the forcing amplitude and $\omega$ is the frequency of the driving. Depending on these parameters, the system exhibits a wide variety of dynamics. Here, we investigate a parameter space region given by $\gamma = 0.15$, $F \in [0.1, 0.5]$ and $\omega \in [0.2, 2.5]$. We search for fractal boundaries in the basins of attraction given by a finite grid in the phase space region $\Omega = \Omega_x \times \Omega_{\dot{x}} = [-2.5, 2.5] \times [-2.5, 2.5]$, with $10^3$ points per axis.

\begin{figure}
  \centering
    \subfloat[\label{fig:duf Sb parameter set}]{\includegraphics[width=0.45\textwidth]{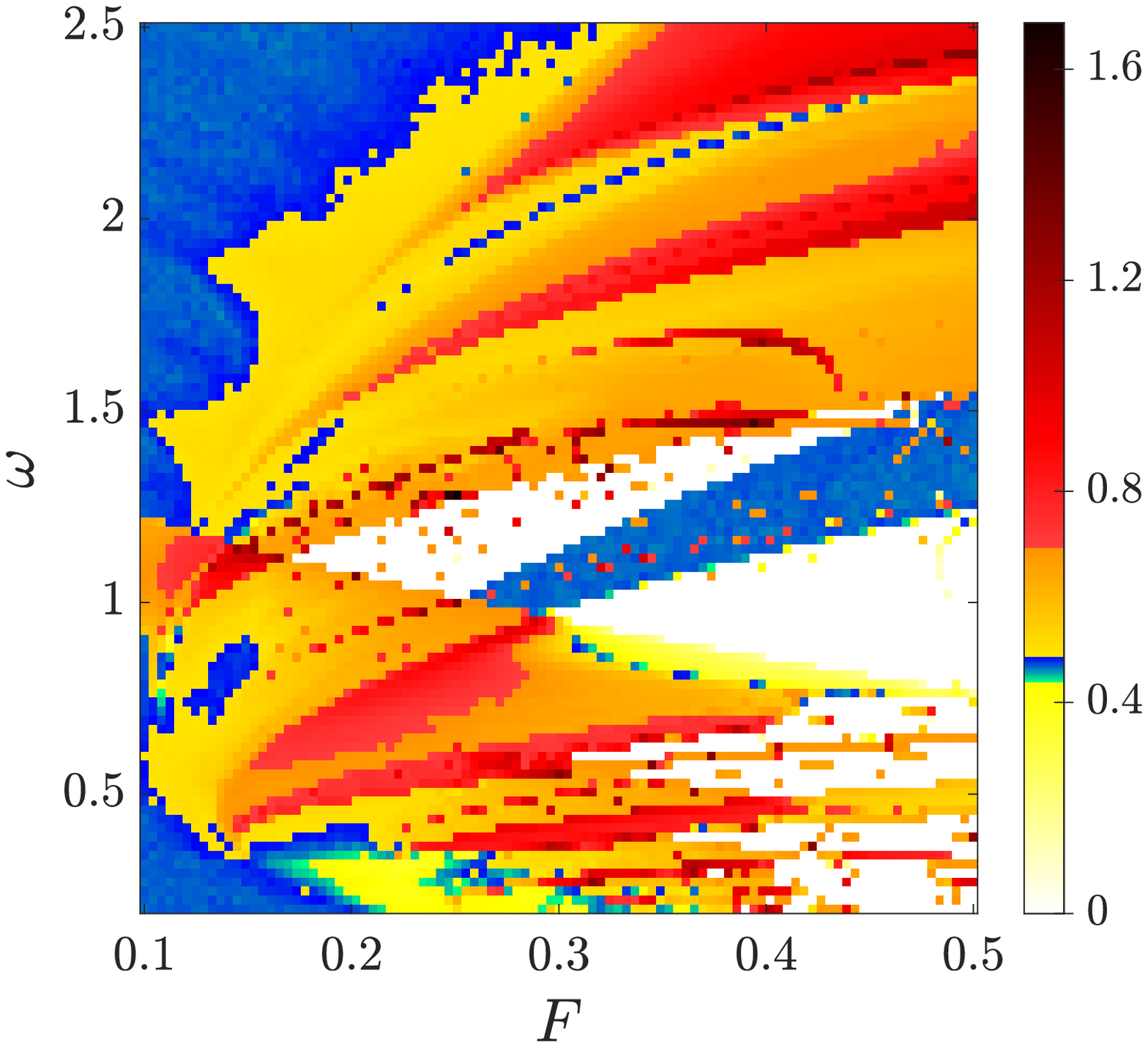}}
    \hspace{0.05\textwidth}
    \subfloat[\label{fig:duf unc exp}]
    {\includegraphics[width=0.45\textwidth]{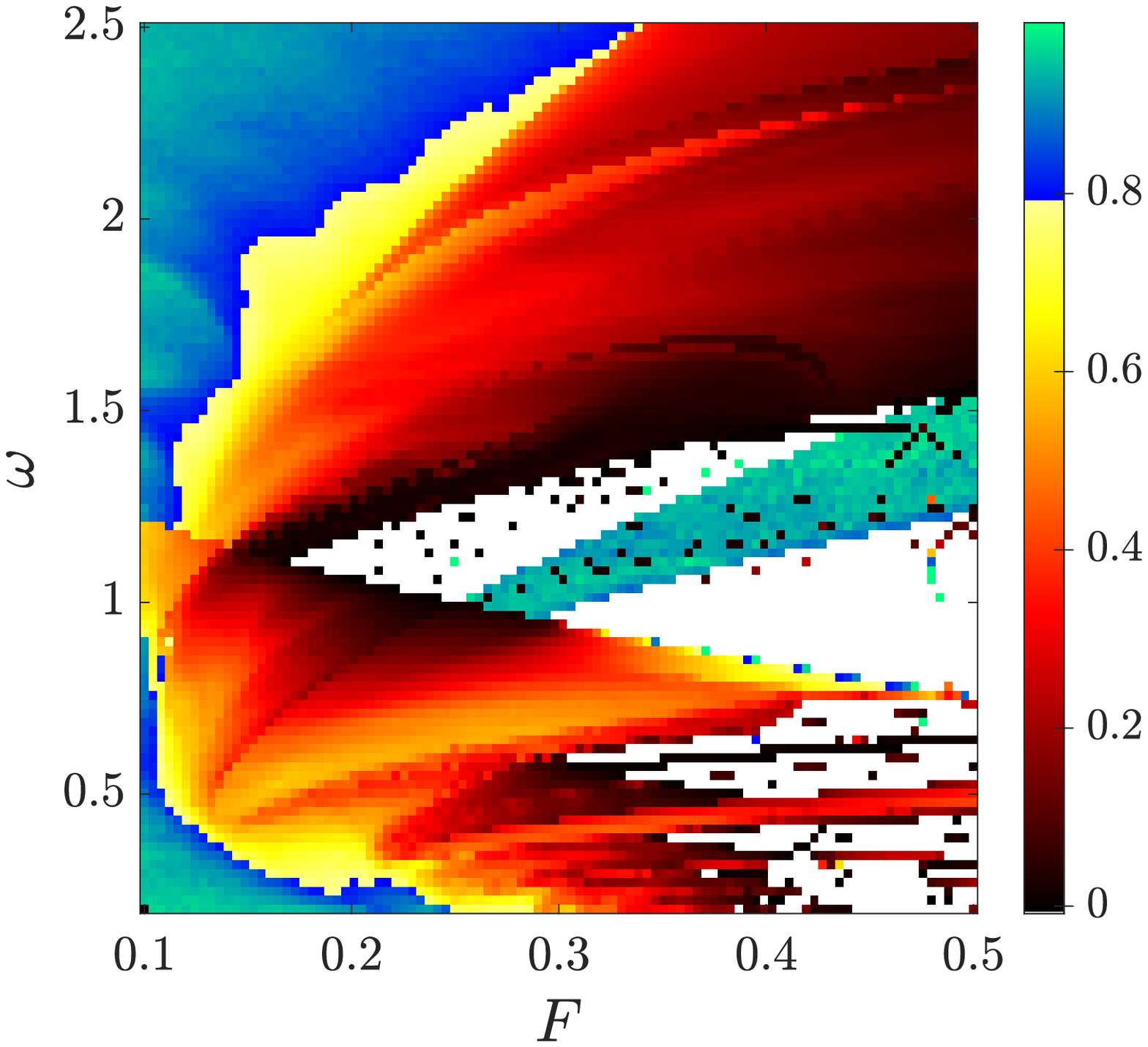}}

  \caption{For the periodically driven Duffing oscillator $\ddot{x} + 0.15 \dot{x} - x + x^3 = F \sin \omega t$ in the parameter space $( F, \omega )$, we obtain similar regions with fractal boundaries (hot colors) and smooth boundaries (cold colors), both for (a) the \textit{$S_{bb}$ fractality test} and (b) the uncertainty exponent $\alpha$. In white, we have regions with a single basin. In (a), the color of each point corresponds to the boundary basin entropy $\hat{S}_{bb}$, for disk boxes with radius $\varepsilon = 0.025$ / $\varepsilon_g = 5$ and $N_b = 10^4$ boxes in the boundary: cold colors are compatible with simple smooth boundaries and hot colors for fractal basin boundaries (red colors are detected by the $\ln 2$ criterion). In (b), we represent the uncertainty exponent with a transition point at $\alpha = 0.8$ from fractal to smooth boundaries; this value is arbitrary and based on the observed basins.}
  \end{figure}

On the following, we use the example to illustrate an \textbf{easy-to-follow recipe for the \textit{$S_{bb}$ fractality test}} for finite grids in two-dimensional phase spaces:
\begin{enumerate}[\bfseries 1.]
 \item \textbf{Choose the box disk radius $\varepsilon$ appropriately}. On the one hand, the smaller the box, the finer scales we can study. On the other hand, we want to have sufficient trajectories per box to get good estimates for the probabilities. For our example, we found that a value of $\varepsilon_g = 5$ / $\varepsilon = 0.025$ gave good results.
 \item \textbf{Verify that we are in the largely extended grid limit, with at least $50 \varepsilon_g$ grid points per axis.} In our example, we had $10^3 = 200 \varepsilon_g$ points per axis.
 \item \textbf{Draw $N_b$ disk boxes in the boundary, uniformly at random from the phase space (and not only at grid points).} In our example, we took $N_b = 10^4$ boxes in the boundary.
 \item \textbf{For all boxes in the boundary, calculate both the observed probabilities of the basins and the observed box's entropy.}
 \item \textbf{Compute the observed boundary basin entropy $\hat{S}_{bb}$ and its standard deviation $\sigma_{\hat{S}_{bb}}$ (Eq.~\ref{eq: Sb statistical error}).}
 \item \textbf{Compute the upper-bound systematic error  $\delta_{\hat{S}_{bb}}^\text{UB}$ (Eq.~\ref{eq:Sbb UB systematic error}) for the current value of $\varepsilon_g$.} In our example, we had $\delta_{\hat{S}_{bb}}^\text{UB} = 0.0474$.
 \item \textbf{Check if the computed value of $\hat{S}_{bb}$ lies in the interval provided by the systematic and statistical errors.} If the value is outside the interval, we can affirm that according to our test the boundary is fractal. Otherwise, it is most likely to be a smooth boundary, although there could be pathological cases leading to wrong results. It is important to recall that this fractality test is a sufficient but not necessary condition.
\end{enumerate}
We display the results in the parameter space of Fig.~\ref{fig:duf Sb parameter set}. The color of each parameter value corresponds to its boundary basin entropy $\hat{S}_{bb}$ value: white is for regions with a single basin, cold colors are for regions compatible with simple smooth boundaries and hot colors are for fractal basin boundaries. In particular, red colors are for fractal basin boundaries detected by the $\ln 2$ criterion. We can observe that this just accounts for a $27\%$ fraction of all the fractal basin boundaries detected by the single-scale fractal basin boundary test.

To evaluate the performance of the method, we compare it to the uncertainty exponent $\alpha$ (Fig.~\ref{fig:duf unc exp}). From the observed basins, we have chosen the arbitrary value of $\alpha = 0.8$ as the transition point between smooth and fractal boundaries: $\alpha \geq 0.8$ correspond to smooth boundaries (cold colors) and $\alpha < 0.8$ to fractal basin boundaries (hot colors).  Again, parameters with a single basin are represented in white. Indeed, both the uncertainty exponent and the \textit{$S_{bb}$ fractality test} give qualitatively similar regions with fractal boundaries. However, there are some quantitative differences due to numerical errors in both methods.

\section{\label{sec:Extension any dimension}Extension to any phase space dimension}
We can generalize the test to phase spaces for any dimension. The test again compares for small box scales $\varepsilon$ the boundary basin entropy $S_{bb}$ of the boundary under study to the theoretical value of that of a flat boundary of the corresponding dimension. Here, we only consider the case with an exact knowledge of the phase space. We believe that the study for basins with finite grids can be developed similarly to the two-dimensional case.

\begin{figure}[h]
  \centering
    \subfloat[\label{fig:D Dim px disk flat}]
    {\includegraphics[width=0.4\textwidth]{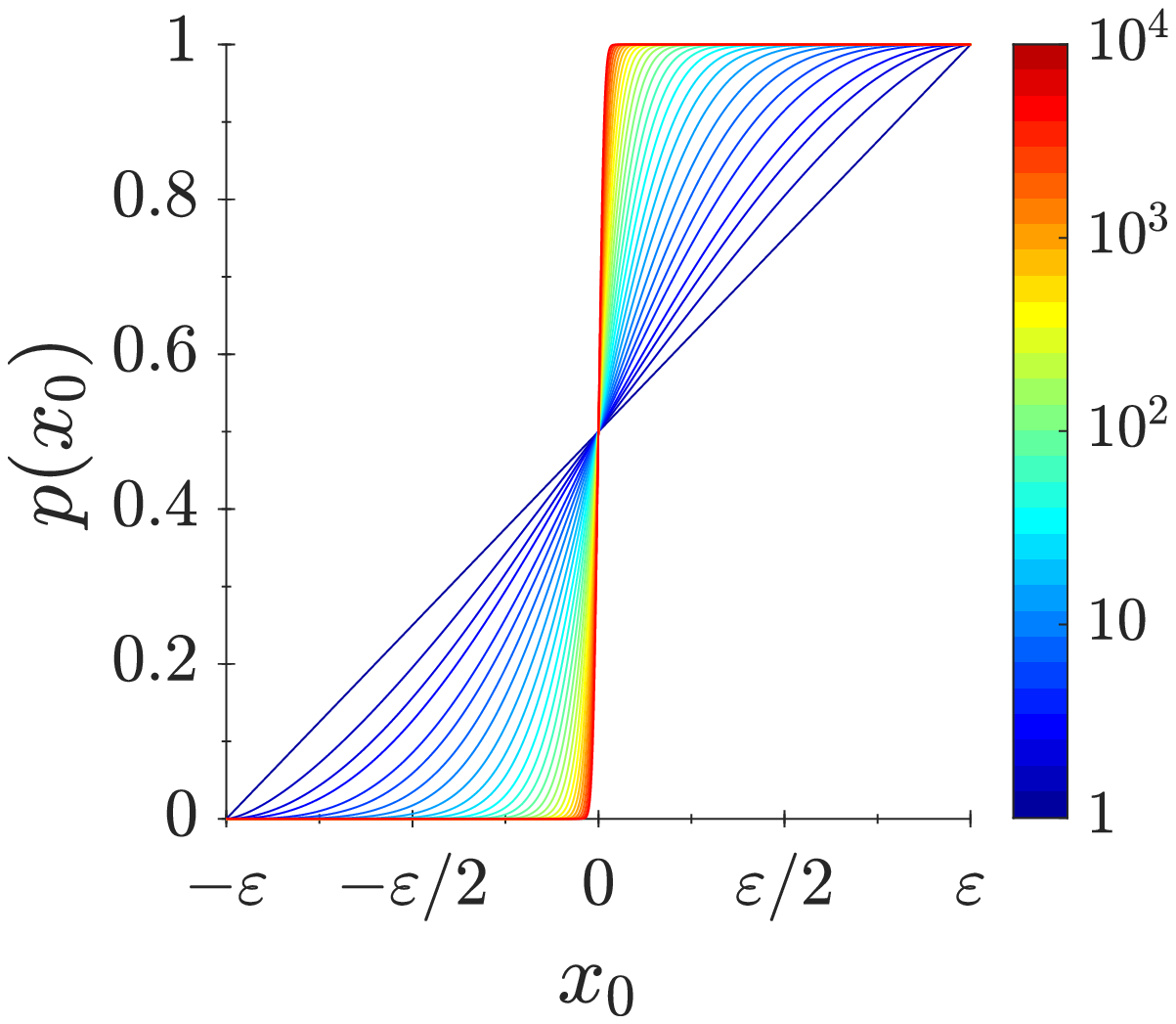}}
       \hspace{0.1\textwidth}
    \subfloat[\label{fig:D Dim Sbb}]{\includegraphics[width=0.4\textwidth]{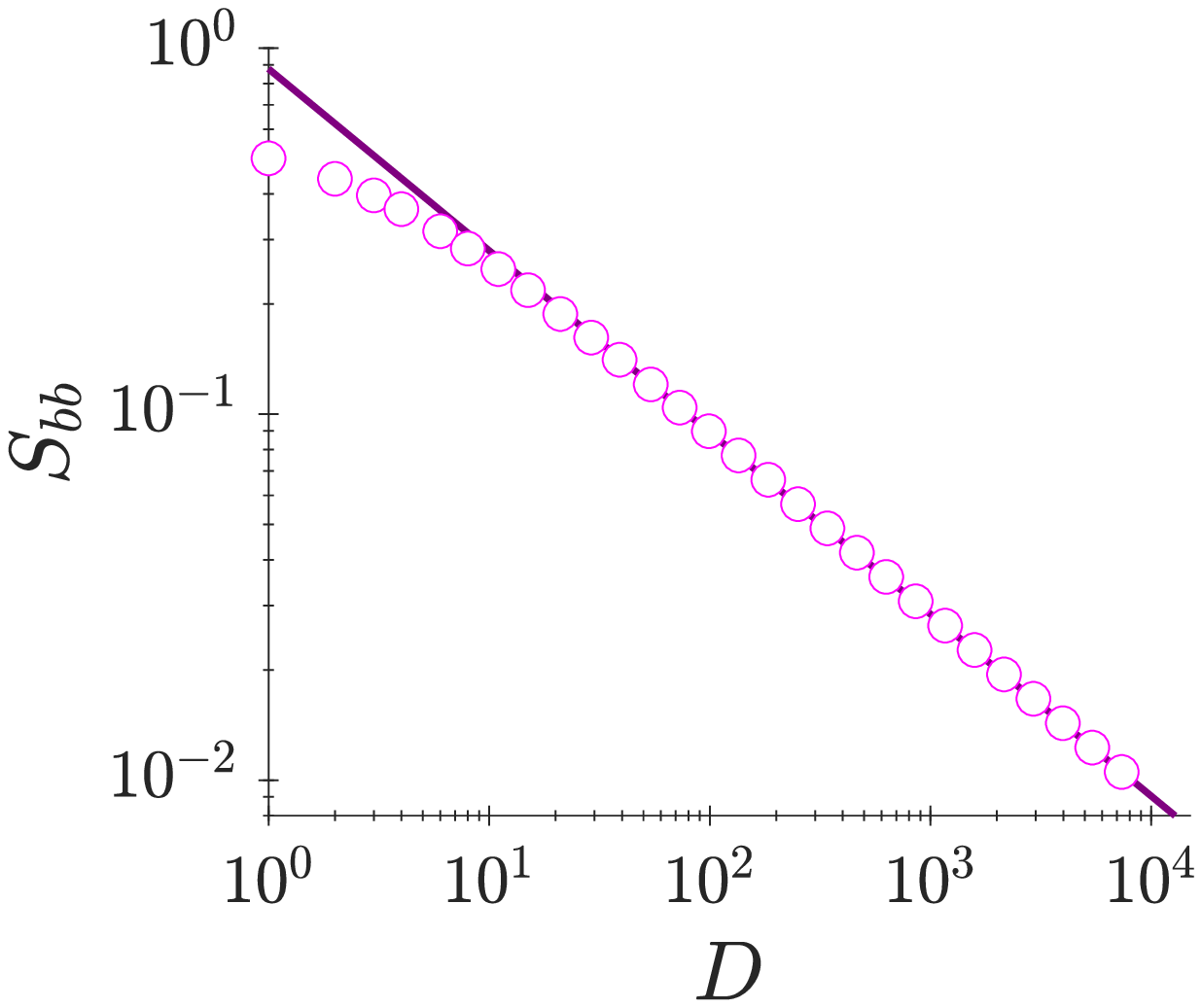}}
  \caption{For a hyperball box and a flat boundary in the phase space, we plot (a) for several dimensions of the phase space (colorbar), the probability $p(x_0)$ to have a hyperball point in one basin in function of the box center coordinate $x_0$, which indicates the distance to the flat boundary; and (b) the boundary basin entropy $S_{bb}$ relationship with the dimension $D$ of the phase space (magenta circles). $S_{bb}$ decreases potentially fast with a power fit $S_{bb} = A D^B$ (purple line): $A=0.898 \pm 0.006$ and $B=-0.4995 \pm 0.0012$.}
  \end{figure}

For a given dimension, we can derive the theoretical value of the boundary basin entropy of a phase space with a flat boundary. The box is now an hyperball of radius $\varepsilon$ centered at a coordinate $x_0$ in the direction that indicates the distance to the flat boundary. Furthermore, the probability $p(x_0)$ to have a hyperball point in one basin is the fraction of the hyperball given by the hyperspherical cap~\cite{li2011concise}:
\begin{equation}
    p(x_0)  = -\frac{1}{2} \text{sgn}\left(\frac{x_0}{\varepsilon} \right) I_{1-\left( \frac{x_0}{\varepsilon} \right)^2} \left( \frac{D+1}{2}, \frac{1}{2} \right) + \Theta \left(\frac{x_0}{\varepsilon}\right),\label{eq:D Dim px0}
\end{equation}
where $D$ is the dimension of the phase space, $\text{sgn}(x)$ is the sign function, $I_x \left(a,b \right)$ is the regularized incomplete beta function and $\Theta(x)$ is the Heaviside function. We have this quantity plotted for several dimensions (see colorbar) in Fig.~\ref{fig:D Dim px disk flat}. For $D=1$, the probability $p(x_0)$ is linear; for larger values of $D$, it deviates from this behavior; and in the limiting case $D \to \infty$, $p(x_0)$ becomes dominated by the $\Theta \left(\frac{x_0}{\varepsilon}\right)$ term and has a switching behavior. This large dimension behavior is understood because most of the volume of a high-dimensional hyperball lies within two parallel hyperplanes at a small distance from its center of order $\mathcal{O} \left( \frac{\varepsilon}{\sqrt{D-1}} \right)$~\cite{hopcroft2012computer}.
Moreover, we calculate the boundary basin entropy $S_{bb}$ analogously to the two-dimensional case, following Eq.~\ref{eq:sb sliding box}. We have the $S_{bb}$ relationship with the dimension $D$ in Fig.~\ref{fig:D Dim Sbb}; we can see how the boundary basin entropy goes to zero in the infinite dimension limit following the expression $S_{bb} \sim D^{-\frac{1}{2}}$. We have tabulated the boundary basin entropy $S_{bb}$ for the first five $D$ dimensions in table~\ref{tab:D dim Sb}.
\begin{table}[h]
\centering
\caption{Boundary basin entropy $S_{bb}$ for hyperball boxes in a flat boundary of a $D$ dimensional phase space, for the first five $D$ dimensions.}\label{tab:D dim Sb}
\begin{tabular}{ l @{\qquad} l }
\hline
     $D$ & $S_{bb}$ \\
     \hline
     1 & 0.499999(9)  \\
     2 & 0.4395093(6) \\
  3 & 0.39609176(4) \\
  4 & 0.36319428(1) \\
  5 &0.33722572 \\
\hline
\end{tabular}
\end{table}

\section{\label{sec:conclusions}Conclusions}

In this paper, we have applied the theory of the basin entropy and related tools to characterize fractal basin boundaries using a single scale. Our work consists on comparing, under a given box scale, the value of the boundary basin entropy $S_{bb}$ of the boundary to the $S_{bb}$ value of a smooth boundary. In contrast to the former basin-entropy-based test, the $\ln 2$ criterion, it achieves both an improved sensitivity and a capacity to identify fractal basin boundaries separating only two basins. We call it \textit{$S_{bb}$ fractality test}. Nonetheless, this test is still a sufficient but not necessary condition. One could think about an improvement by means of the probability density function (PDF) of the probabilities in the boundary boxes. The configuration of a PDF is more restrictive than an average (the boundary basin entropy), but it is also more demanding to implement.

On the other hand, testing for fractal basin boundaries with this line of work has important advantages compared to the classical approach given by the uncertainty exponent $\alpha$. While both methods hold numerical limitations in the infinitely fine scale, the uncertainty exponent requires accessing multiple scales of the system. This is not only numerically inconvenient but, for some physical systems, it may be impossible. Indeed, our methods only require accessing a single scale and have a natural formulation for experimental basins of attraction.

Finally, we believe the extension for basins in any phase space dimension could prove useful for studying the unpredictability in higher-dimensional systems, a generally unexplored area in the field; in particular, it could complement current studies for network systems~\cite{menck2013basin,menck2014dead,rakshit2017basin}.

\section*{Acknowledgements}
This work has been supported by the Spanish State Research Agency (AEI) and the European Regional Development Fund (ERDF, EU) under Projects No.~FIS2016-76883-P and No.~PID2019-105554GB-I00.





\end{document}